\documentclass[a4paper]{report}
\usepackage[utf8]{inputenc}
\usepackage[T1]{fontenc}
\usepackage{RJournal}
\usepackage{amsmath,amssymb,array}
\usepackage{booktabs}

\begin{document}

\sectionhead{Graphical outputs and Spatial Cross-validation for the \textbf{R-INLA} package using \textbf{INLAutils}}
\volume{XX}
\volnumber{YY}
\year{20ZZ}
\month{AAAA}

\begin{article}
\title{Graphical outputs and Spatial Cross-validation for the \textbf{R-INLA} package using \textbf{INLAutils}}
\author{by Tim C.D. Lucas, Andre Python and David W. Redding}

\maketitle

\abstract{
Statistical analyses proceed by an iterative process of model fitting and checking.
The \textbf{R-INLA} package facilitates this iteration by fitting many Bayesian models much faster than alternative MCMC approaches. As the interpretation of results and model objects from Bayesian analyses can be complex, the \texttt{R} package \textbf{INLAutils} provides users with easily accessible, clear and customisable graphical summaries of model outputs from \textbf{R-INLA}. Furthermore, it offers a function for performing and visualizing the results of a spatial leave-one-out cross-validation (SLOOCV) approach that can be applied to compare the predictive performance of multiple spatial models. In this paper, we describe and illustrate the use of (1) graphical summary plotting functions and (2) the SLOOCV approach. We conclude the paper by identifying the limits of our approach and discuss future potential improvements.
}

\section{Introduction}

From its inception in 2009, the \texttt{R} \citep{R} package \textbf{R-INLA} \citep{Rue2009,Martins2013} has provided a convenient framework to fit latent Gaussian models within a Bayesian framework using R commands. Latent Gaussian models represent a wide and flexible class of models that includes mixed effects models and spatial and spatio-temporal models that can be applied to areal, geostatistical, and point process data. For models with a continuous spatial component, the approach combines the integrated nested Laplace approximation (INLA) \citep{Rue2009} with the stochastic partial differential equation approach (SPDE) developed by \citet{Lindgren2011}. 
The INLA/SPDE approach represents a computationally efficient alternative to Markov Chain Monte Carlo (MCMC) \citep{Lindgren2015}. Despite MCMC's high flexibility on both data and model types, estimating the posterior distribution for the parameters remains computationally expensive in this framework. 
The superior computational performance of INLA/SPDE allows fitting a broad range of models applied to large data and running systematic cross-validation processes which would require considerably more time with MCMC algorithms.

Users who are not familiar with the usage of this statistical software might face difficulties in extracting and visualizing summary characteristics of the model outputs generated by \textbf{R-INLA}. Outputs from \textbf{R-INLA} consist of large nested lists, whose elements can be difficult to be identified. As a fundamental part of iterative model building, it is important for summaries and visualisations of models to be as easily accessible as possible \citep{gabry2017visualization}. 
In addition, cross-validation of spatial models as currently implemented in \textbf{R-INLA} does not account for spatial autocorrelation in the data. 
Predictions made geographically far away from the data will be less good than predictions made near the data as they rely solely on covariates without any contribution from the spatial random field.
Spatial leave-one-out cross-validation is one solution to this problem where data near the hold-out data point is also removed.
Here we present functions from the package \textbf{INLAutils} that provides functions for streamlining these aspects of the Bayesian modelling process with  \textbf{R-INLA}.

Section~\ref{section:graph} illustrates the use of commands to generate graphical summaries of \textbf{R-INLA} model outputs. Section~\ref{section:sloo} illustrates the command used to perform and visualize the results of a SLOOCV procedure applied to multiple models. In Section~\ref{section:conclusion} we discuss the limits of the package capabilities and identify potential future improvements.   

\section{Installation}

Due to technical issues with building the C binaries needed by \textbf{R-INLA}, the package is not available from CRAN. It can be installed the following command

\begin{example}
install.packages("INLA", 
                 repos = c(getOption("repos"), 
                           INLA = "https://inla.r-inla-download.org/R/stable"), 
                 dep = TRUE)
\end{example}
In turn, this means \textbf{INLAutils} is not on CRAN. It is available from \url{www.github.com/timcdlucas/INLAutils} or can be installed with the following command.
\begin{example}
devtools::install_github('timcdlucas/INLAutils')
\end{example}

\section{Example data analysis} \label{s:eg}

To illustrate the various aspects of this package we will use the Meuse dataset \citep{rikken1993soil}  that is available in the \texttt{sp} package \citep{sp}. 
This dataset contains 155 observations of concentrations of four heavy metals.
We fit a latent Gaussian regression model \texttt{modform} on a normal response (cadmium concentration), which includes an intercept, \texttt{y.intercept}, three covariates (\texttt{elev}, \texttt{dist}, \texttt{om}), a spatial component, \texttt{f(spatial.field, model = spde)}, and an unstructured Gaussian error term (not displayed in the formula). 
In \textbf{R-INLA}, the spatial component is represented by a Gaussian Markov Random Field, with a Matérn covariance function. 
The flexibility of the the Gaussian Markov Random Field is controlled by two hyperparameters, $\theta_1$ and $\theta_2$.
For comparison purposes, we fit the same model without the spatial component, \texttt{modform2}. 
We show below the data preparation and code used to run the models. 
Further details on coding spatial models with \textbf{R-INLA} can be found in \citet{blangiardo2015spatial}.

\begin{example}
require(INLAutils)
require(INLA)
require(sp)

data(meuse)

# Define the models
modform <- cadmium ~ -1 + y.intercept + elev + dist + om + 
             f(spatial.field, model = spde)
modform2 <- cadmium ~ -1 + y.intercept + elev + dist + om

# Scale the spatial coordinates to make mesh construction easier
coords <- scale(meuse[, c('x', 'y')])
colnames(coords) <- c('long', 'lat')
dataf1 <- sp::SpatialPointsDataFrame(coords = coords, data = meuse[, -c(1:2)])

mesh <- inla.mesh.2d(loc = sp::coordinates(dataf1), 
                     max.edge = c(0.2, 0.5), cutoff = 0.1)
spde <- inla.spde2.matern(mesh, alpha=2) # SPDE model is defined
A <- inla.spde.make.A(mesh, loc = sp::coordinates(dataf1)) # projector matrix
dataframe <- data.frame(dataf1) # get dataframe with response and covariate

# make index for spatial field
s.index <- inla.spde.make.index(name="spatial.field",n.spde=spde$n.spde)

# Prepare the data
stk <- inla.stack(data=list(cadmium=dataframe$cadmium),
                      A=list(A,1), 
                      effects=list(c(s.index,list(y.intercept=1)),
                                   list(dataframe[, 5:7])),
                      tag='est')

out <- inla(modform, family = 'normal',
            data = inla.stack.data(stk, spde = spde),
            control.predictor = list(A = inla.stack.A(stk), link = 1),
            control.compute = list(config = TRUE), 
            control.inla = list(int.strategy = 'eb'))

\end{example}

\section{Graphical Summaries with the \texttt{autoplot}, \texttt{ggplot\textunderscore inla\textunderscore residuals} and \\ \texttt{ggplot\textunderscore projection\textunderscore shapefile} functions}
 \label{section:graph}

\subsection{The \texttt{autoplot.inla} command}
\label{subsection:autoplot}

\begin{figure}[htbp]
  \centering
  \includegraphics[width = \textwidth]{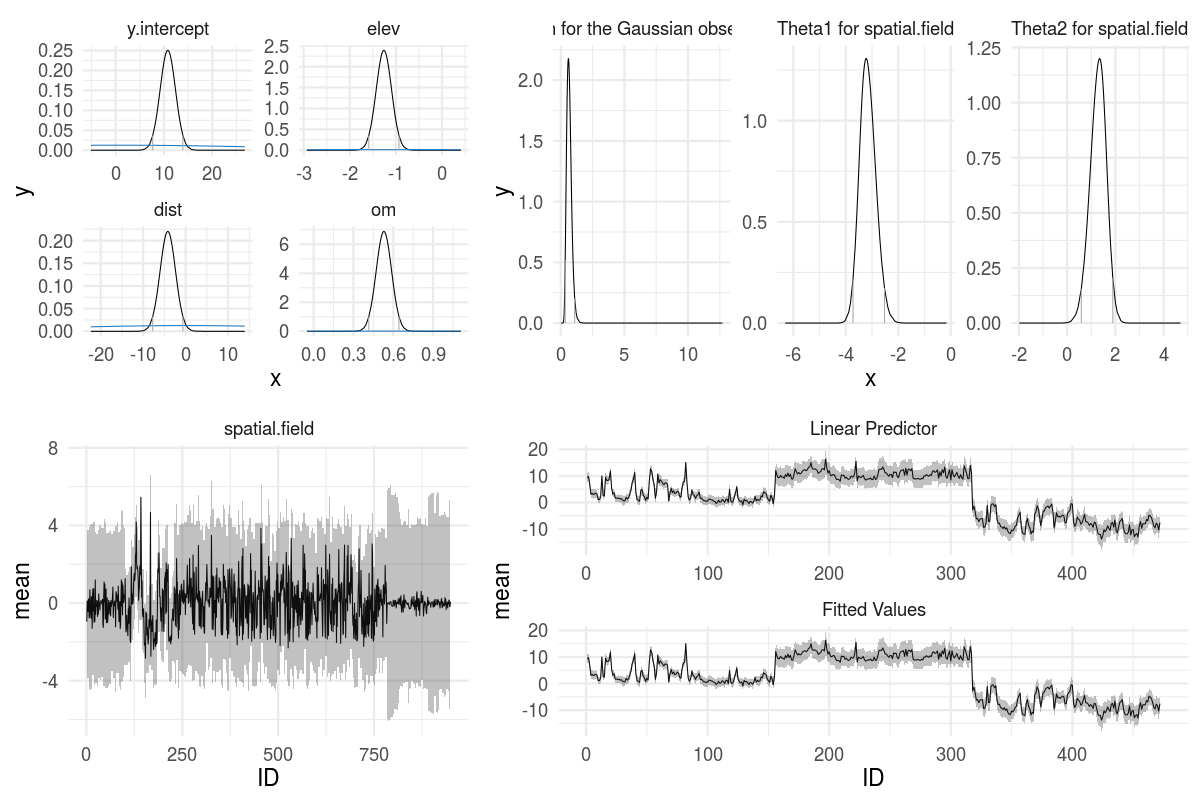}
  \caption{ (\textit{four plots top-left}) Marginal posterior distributions (black lines), prior distributions (blue lines) and 95\% credible interval (grey vertical lines) of \texttt{y.intercept} and covariate coefficients for \texttt{elev}, \texttt{dist} and \texttt{om}. (\textit{three plots top-right}) Marginal posterior distributions (black lines) and 95\% credible interval (grey vertical lines) of the precision of the Gaussian unstructured error term and two hyperparameters for the spatial random field. (\textit{bottom-left}) Marginal posterior mean and 95\% credible intervals of the spatial random effects. (\textit{two plots bottom-right}) Marginal posterior mean and 95\% credible intervals of the linear predictor,  and fitted values.}
  \label{figure:autoplot1}
\end{figure}

We provide methods for the \texttt{autoplot} command from \texttt{ggplot2} for \textbf{R-INLA} models and meshes. Using \texttt{ggplot2} means all visual aspects can be easily fine-tuned.
The \texttt{autoplot.inla} method allows the user to visualize outputs from any \textbf{R-INLA} object (Figure~\ref{figure:autoplot1}). 
The command reimplements much of the functionality from the \textbf{R-INLA} \texttt{plot} command but uses \texttt{ggplot2} as the plotting backend. 
By default the function plots provides a visual overview of (\textit{top-left}): the marginal posterior distributions, prior distributions and credible intervals of the intercept (\texttt{y.intercept}) and coefficients for the covariates (\texttt{elev, dist, om}); (\textit{top-right}): the marginal posterior distributions and credible intervals of model hyperparameters; the marginal posterior mean and 95\% credible intervals of any random effects if specified in the model and; (\textit{bottom-left}) and the linear predictor and fitted values (\textit{bottom-right}). 
As the link function is the identity function the linear predictor and fitted values are the same in this case.
These subplots can be activated or deactivated with the \texttt{which} argument.
This function returns a list of \texttt{ggplot2} plots which can be plotted together using \texttt{plot\textunderscore grid} from the \texttt{cowplot} package \citep{cowplot}.
The command can be used with any \textbf{R-INLA} model object though subplots that plot random effects are only created for models with random effects.
Which subplots to create can be selected with the \texttt{which} argument.
The other arguments to the command are \texttt{priors} which determines whether the priors of the fixed effects are plotted and \texttt{CI} which determines whether 95\% credible intervals are plotted for both the fixed effects and hyperparameters.

\begin{example}
p <- autoplot(out)
cowplot::plot_grid(plotlist = p)
\end{example}

\begin{figure}[htbp]
	\centering
	\includegraphics[width = 0.7\textwidth]{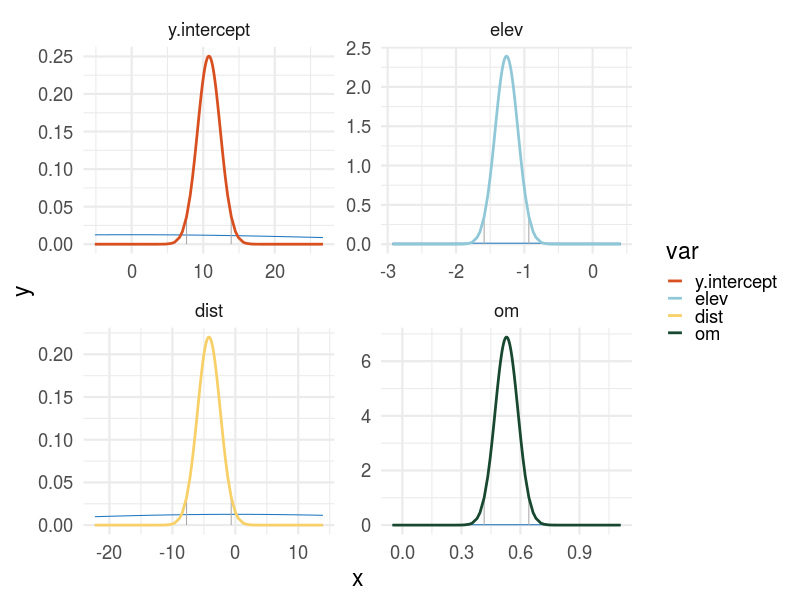}
	\caption{Visualization of the posterior distribution of the intercept (\texttt{y.intercept}) and covariates' coefficients using thick lines with four distinct colours: \texttt{y.intercept} (red), \texttt{elev} (light blue), \texttt{dist} (orange), and \texttt{om} (green). The prior distributions (blue lines) and 95\% credible intervals (grey vertical lines) are also shown.}
	\label{figure:autoplot2}
\end{figure}

In the following example (Figure~\ref{figure:autoplot2}), we use \texttt{ggplot2} to modify the plot of the the posterior distributions of the intercept and coefficients for the covariates, by selecting the first object of the \texttt{autoplot} output, with the command \texttt{p[[1]]}. The size and colour of the lines are modified using the \texttt{ggplot2} syntax with a convenient colour palette \citep{palettetown}.
  
\begin{example}
require(ggplot2)
require(palettetown) # a convenient palette

p[[1]] + 
  geom_line(aes(colour = var), size = 1.3) +
  palettetown::scale_colour_poke(pokemon = 'Charmander', spread = 4)
\end{example}

\subsection{The \texttt{ggplot\textunderscore inla\textunderscore residuals} command}

Other important summary characteristics of \textbf{R-INLA} outputs can be visualized. Here, the \texttt{ggplot\textunderscore inla\textunderscore residuals} command is applied to the results of the \textbf{R-INLA} outputs and the response data, extracted from the \texttt{meuse} object, as illustrated below:

\begin{example}
ggplot_inla_residuals(out, meuse$cadmium, binwidth = 0.1)
\end{example}

\begin{figure}[htbp]
	\centering
	\includegraphics[width = 0.7\textwidth]{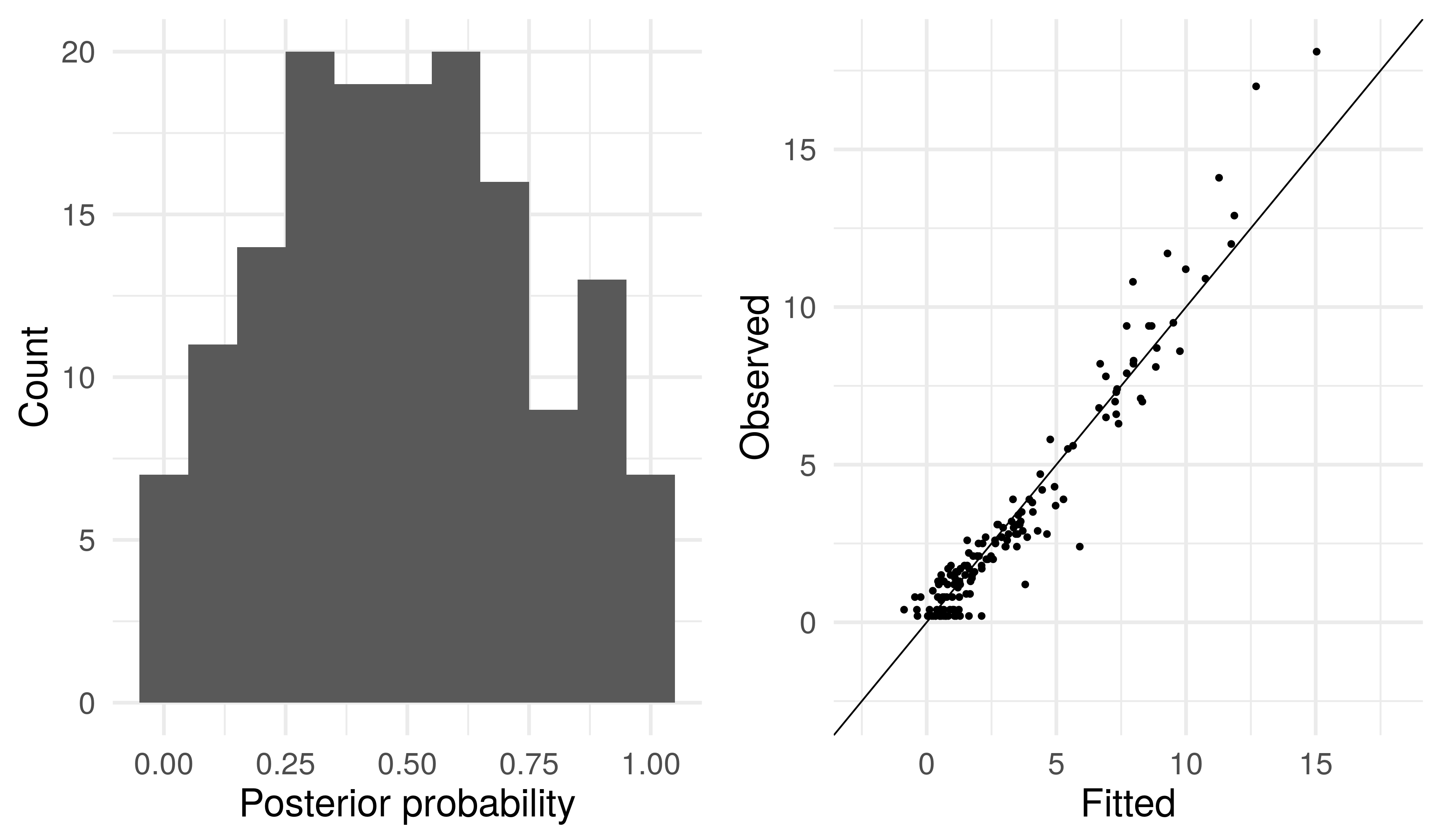}
	\caption{(\textit{Left}): histogram of the posterior probability of a replicate of each observation, and (\textit{right}): relationship between the observed (x-axis) and the predicted (y-axis) values. The identity function $y=x$ is illustrated by the black line.}
	\label{figure:residual}
\end{figure}

This function plots an histogram of the posterior probability of a replicate of each observation (Figure~\ref{figure:residual} \textit{left}).
The binwidth can be controlled with the argument \texttt{binwidth}. 
If the model is well calibrated (i.e.\thinspace if the uncertainty estimates are accurate) the posterior probabilities are expected to be uniformly distributed (and therefore the height of the bins should be similar).
In the example, the humped distribution suggests that the model is underconfident in its predictions. 
Figure~\ref{figure:residual} (\textit{right}) illustrates the relationship between the observed (x-axis) and the predicted (y-axis) values. 
Values above and below the line (identity function: $y=x$) correspond to overestimation and underestimation of the data by the model, respectively.  
The two plots are returned as a list of \texttt{ggplot} objects.

\subsection{The \texttt{autoplot.inla.mesh} command}

For continuous-space models in \textbf{R-INLA}, users need to define a mesh on top of which the stochastic partial differential equation (SPDE) is built (see \citet{Lindgren2015} for further details). In a two-dimensional spatial model, the mesh consists of triangles that can be defined via several parameters, as shown in Section \ref{s:eg}. The returned mesh object has the class \texttt{inla.mesh} and we provide a seperate \texttt{autoplot} method for objects of this class which can be used as follows:

\begin{example}
autoplot(mesh)
\end{example}

\begin{figure}[htbp]
	\centering
	\includegraphics[width = 0.7 \textwidth]{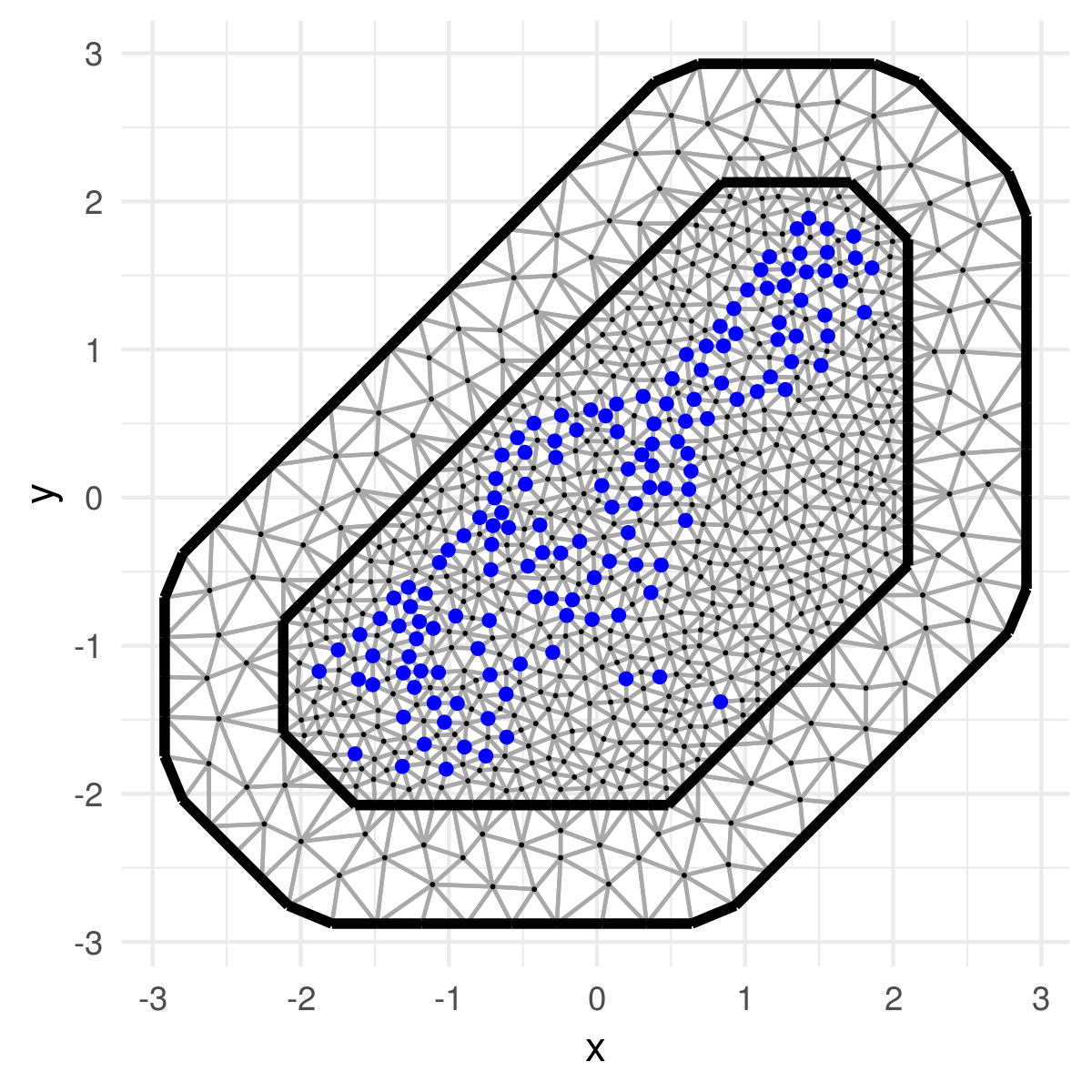}
	\caption{The SPDE mesh on top of which the stochastic partial differential equation (SPDE) is built and locations of the observations (\textit{blue dots}). The coordinates represent the Rijksdriehoek (RDH) (Netherlands topographical) that have then been scaled and centered.} 
	\label{figure:mesh}
\end{figure}

When applied to a mesh object, the \texttt{autoplot} command provides an elegant visualization of the mesh and the location of the observations. The values of the axes refer to the coordinate system associated with the observations (Figure~\ref{figure:mesh}).

\subsection{The  \texttt{ggplot\textunderscore projection\textunderscore shapefile} command}

\begin{figure}[htbp]
	\centering
	\includegraphics[height = 0.7 \textwidth]{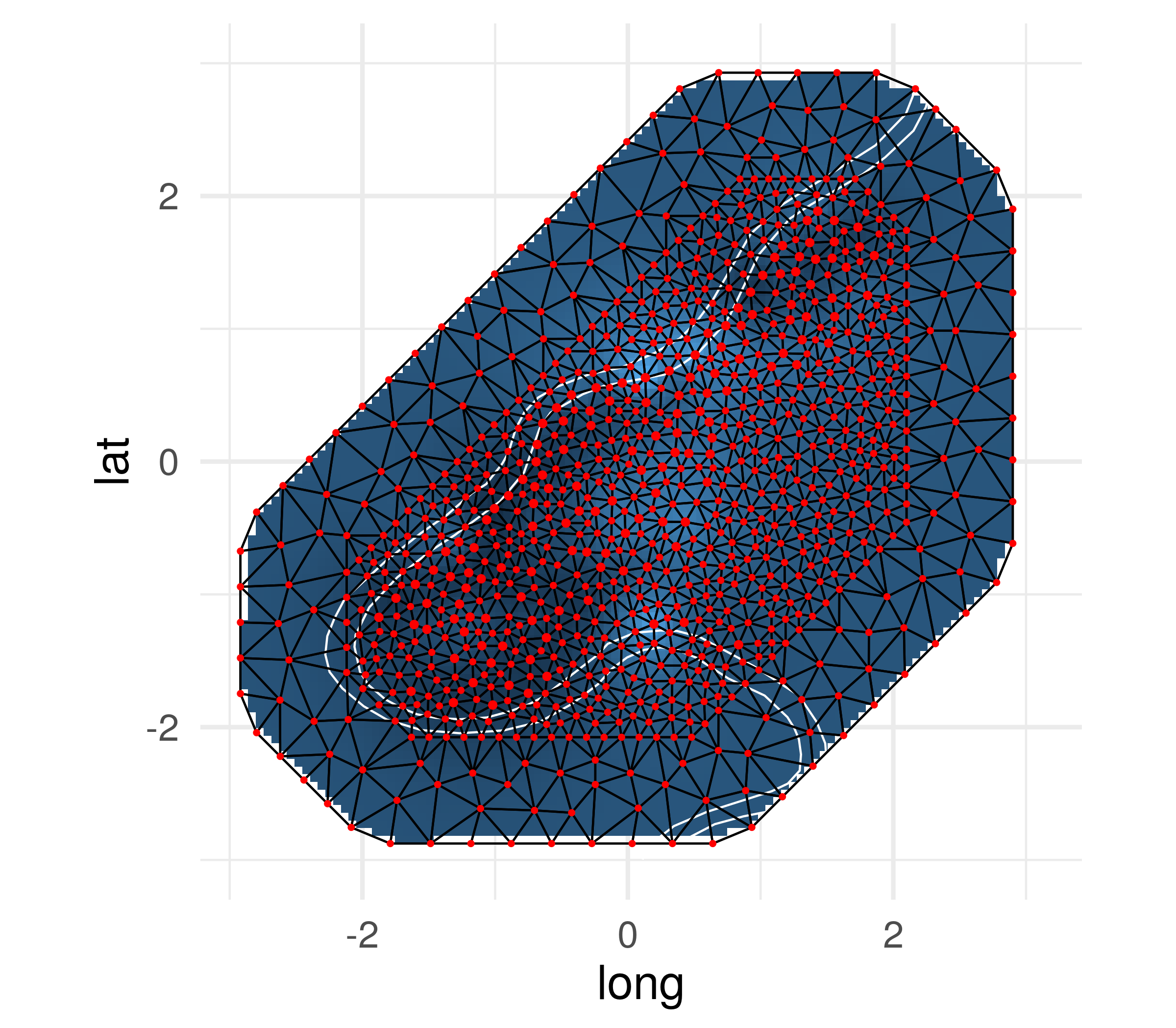}
	\caption{Visualization of the mesh with vertices (\textit{red points}) and edges \textit{black triangles}, study area (\textit{polygon with white border}) and values of the Gaussian random field (\textit{blue shade}). The coordinates represent the Rijksdriehoek (RDH) (Netherlands topographical) that have then been scaled and centered.}
	\label{figure:meshandfield}
\end{figure}

It is convenient to visualize together the study area, mesh, and values of the random field. The corresponding objects are often of different classes. 
Study areas are commonly represented by spatial polygons (representing geographically relevant areas such as countries or in this case the Meuse river), while the mesh and the random field are distinct \textbf{R-INLA} objects. 
As illustrated in Figure~\ref{figure:meshandfield}, the \texttt{ggplot\textunderscore projection\textunderscore shapefile} command allows the user to visualize the study area, mesh, and values of the random field all together. 
This function takes as its first argument either a \texttt{RasterLayer} or a matrix. 
If a matrix is provided, an \texttt{inla.mesh.projector} object must also be provided as the second argument.
An optional \texttt{SpatialPolygonsDataFrame} and \texttt{inla.mesh} can also then be provided as arguments \texttt{spatialpolygons} and \texttt{mesh} respectively.
The function returns a \texttt{ggplot} object which can then be altered as usual.

\begin{example}

data(meuse.riv)
projector <- inla.mesh.projector(mesh)
projection <- inla.mesh.project(projector, 
                                out$summary.random$spatial.field$mean)

# And a shape file and scale using previous scaling values

meuse.riv <- sweep(meuse.riv, 2, attr(coords, "scaled:center"), FUN = "-")
meuse.riv <- sweep(meuse.riv, 2, attr(coords, "scaled:scale"), FUN = "/")
meuse.sr = SpatialPolygons(
             list(Polygons(list(Polygon(meuse.riv)), "meuse.riv"))
           )

# plot
pp <- ggplot_projection_shapefile(projection, projector, meuse.sr, mesh)
pp + coord_equal() + ylim(-3, 3)

\end{example}

\section{Spatial Leave-one-out Cross-validation}
 \label{section:sloo}

It is common to split data into training and validation sets, especially with predictive modelling purposes applied in ecological contexts \citep{hastie2009}. 
A methodologically sound splitting design requires that training and validation data are independent. 
In the presence of spatial autocorrelation, when data held-out for validation are close in space to the training data, the independence between training and validation data can be compromised \citep{dormann2007, hastie2009} and any model selection process may favour overly complex models with potential underestimation of the prediction errors \citep{mosteller1977, roberts2017}.
The risk of overfitting is high in flexible models that include a Gaussian spatial random component.
Furthermore, the risk of under-estimating model error is particularly high when training and validation datasets are close in space but predictions are being made far from the training data.    

In order to mitigate overfitting, generate more reliable error estimates and better select predictive spatial models, a common cross-validation strategy is to split data into spatial blocks \citep{roberts2017}. 
Blocks consist of units of geographical area (e.g. rectangles, hexagons, disks). 
Block cross-validation approaches tend to provide better estimates of the errors in predictions compared to random data splits \citep{burman1994, bahn2013, radosavljevic2014}. However, if spatial blocks follow environmental gradients, for example, gradients of heavy metal concentration following a river flow, predictions can be made outside previously known combinations of predictor values from those learned from the training folds, and hence, lead to extrapolation between cross-validation folds \citep{snee1977, zurell2012}.

The spatial leave-one-out cross-validation (SLOOCV) allows for clear spatial separation between the training and the validation (left-out point) folds \citep{LeRest2014, pohjankukka2017}. 
However, users need to be cautious about defining the radius of the buffer surrounding the hold-out point. 
If the radius is too large, it can produce more similar training sets than blocking strategies using rectangular shapes \citep{hastie2009} while if it is too small, the autocorrelation in the data can still be fitted by the spatial random field.

\begin{figure}[htbp]
\centering
\includegraphics[scale=0.7]{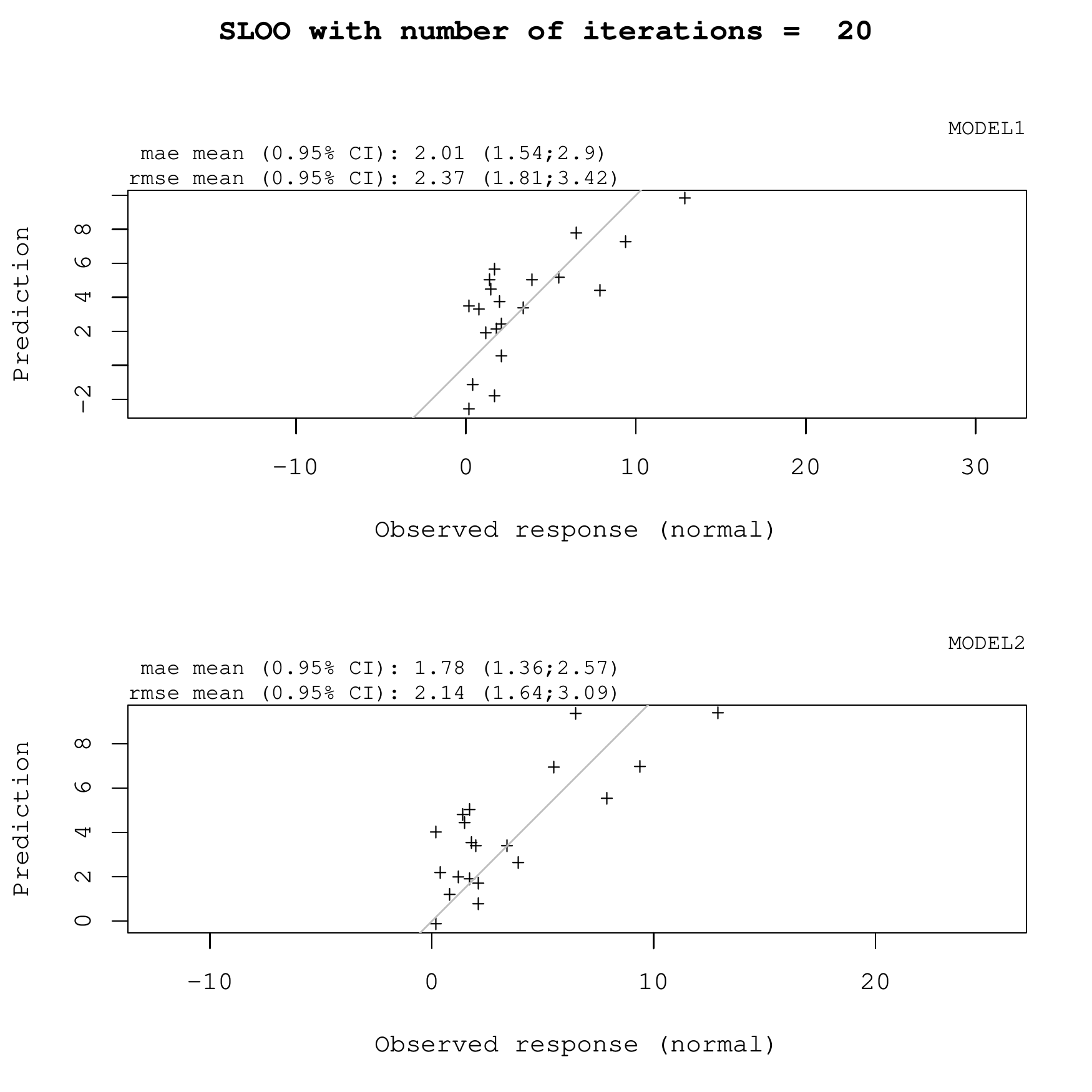}
\caption{Results of the \texttt{inlasloo} procedure, which illustrate the relationship between the observed response (x-axis) and the corresponding predictions (y-axis) for the full model and covariates only model. The identity function ($y=x$) is indicated by a line. \textit{Top-left}: predictive scores (mean and 95\% confidence interval) of the mean absolute error (\texttt{mae}) and root-mean-square error (\texttt{rmse}). \textit{Top-right}: name of the model. The likelihood is a Gaussian distribution (\texttt{normal}), which is indicated in the x-axis label.}
\label{figure:sloo1}
\end{figure}

\begin{figure}[htbp]
\centering
\includegraphics[scale=0.6]{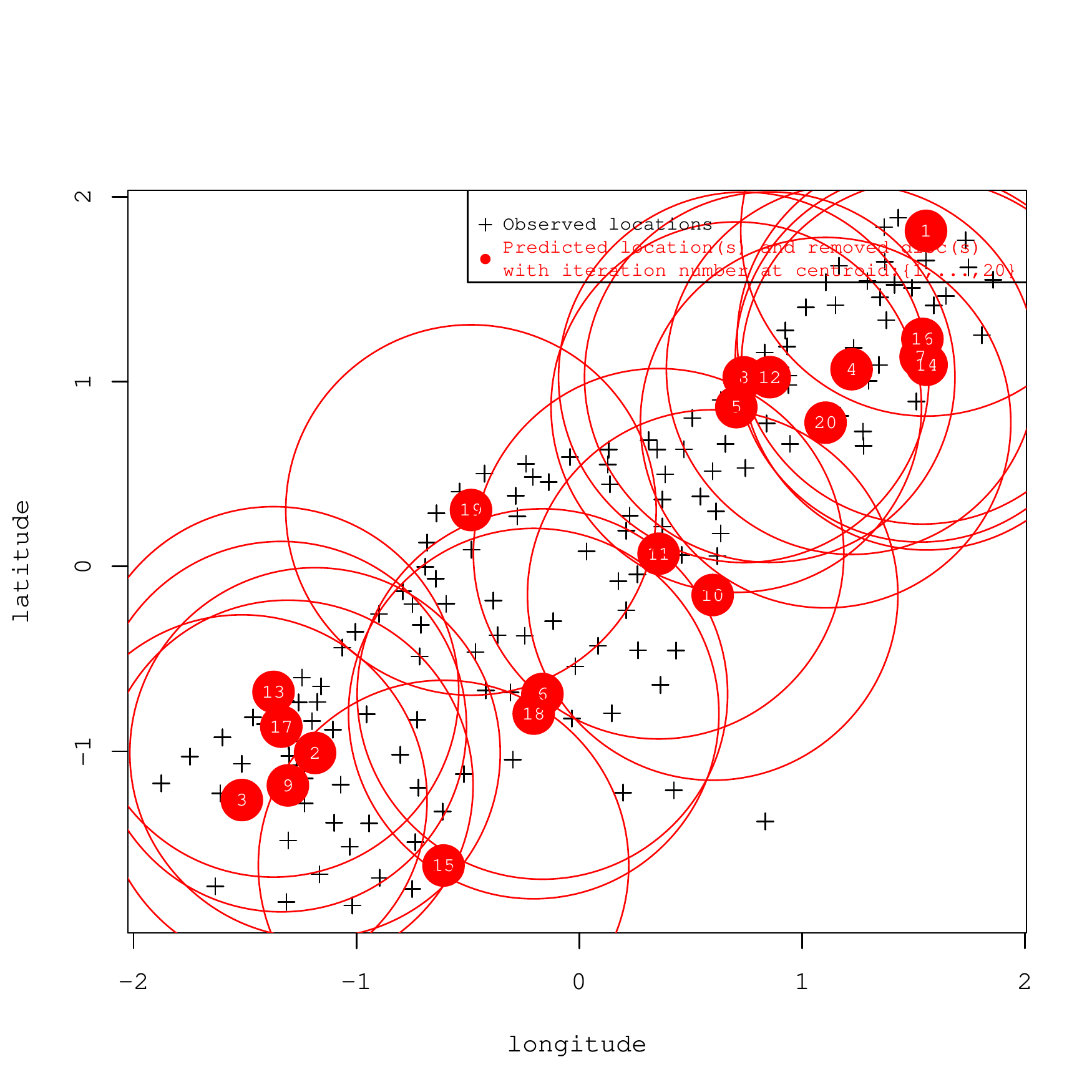}
\caption{Visualization of the left-out observations with their corresponding iteration number (\textit{white number in red disks}), radius defined by the user (\textit{red circles}) and the locations of the remaining data (\textit{black crosses}).}
\label{figure:sloo2}
\end{figure}

\citet{LeRest2014} suggested an SLOOCV approach for GLMs without spatial components fitted to spatial data.
We extend this approach to explicitly spatial models fitted with \textbf{R-INLA}. 
For general information on the approach, please refer to \citet{LeRest2014}. 
Our suggested method (code provided below) follows three steps:
\begin{enumerate}
	\item Remove one observation (validation set) from the initial dataset
  \item Remove observations within a radius (defined by the user), so that the training set is composed of all remaining observations
	\item Predict at the location of the removed observation using parameters estimated with the training set
\end{enumerate}
Steps (1) to (3) are repeated $k$ times.
$k$ is defined by the user and can be less than or equal to the number of data points, $n$, with $k = n$ being the case where all data points are left out once.
The results are illustrated through two plots. 
The first plot (Figure~\ref{figure:sloo1}) highlights the relationship between the predicted and observed values as well as displaying the mean absolute error, $\text{MAE}=\frac{1}{n}\sum_{i=1}^n \lvert y_{observed} - y_{predicted} \rvert$, and the root mean squared error, $\text{RMSE}=\frac{1}{n}\sum_{i=1}^n (y_{observed} - y_{predicted})^2$. 
The second plot displays the location and iteration number corresponding to the left-out observation(s) and corresponding surrounding disk (where observations are removed from the training data) (Figure~\ref{figure:sloo2}).
The locations of the remaining observations, that are not used as test locations are also shown so the user can see which data points will be removed in each model iteration i.e.\thinspace all points within each surrounding disk will be removed for that model fit.

The command \texttt{inlasloo} (R code below) performs the iterative process described above. Note that the procedure performing 20 iterations on the Meuse dataset takes approximately 3-5 minutes to run on a 64Gb RAM Intel Xeon machine. It requires the user to specify the following arguments: 
\begin{itemize}
\item \texttt{dataframe}: A data frame which should contain observations (rows) for each variable of interest (columns). 
\item The names used to defined the variables of interest are defined as strings: response \texttt{y}, geographic coordinates \texttt{long,lat}. 
\item The radius \texttt{rad} (radial distance within which observations are removed from the training set during the procedure) should be given in the unit associated with the coordinates system of the observations. 
\item The formula (\texttt{modform}) should be a formula object or a list of formula objects for testing multiple models. 
\item The mesh object (\texttt{mesh}).
\item RMSE is calculated by default. The \texttt{mae} argument can be set to \texttt{TRUE} if MAE is also required.
\item One or more likelihood families (\texttt{bernoulli, binomial, normal}, etc.) should also be set. Multiple families can be used to compare several models simultaneously, e.g. \texttt{family = c(\textquotesingle normal\textquotesingle,\textquotesingle bernoulli\textquotesingle}).
\end{itemize}

In order to illustrate how SLOOCV can be used to test predictive performance we compare two models fitted to the Meuse dataset. 
We fit the model with three covariates and a spatial random field  as used above (\texttt{modform}) and compare it to a model with only covariates (\texttt{modform2}).
In this case, there is little evidence that one model has better predictive performance than the other as for both MAE and RMSE the confidence intervals overlap (Figure~\ref{figure:sloo1}).

\begin{example}

out.field <- inla.spde2.result(out,'spatial.field', spde, do.transf = TRUE)
range.out <- inla.emarginal(function(x) x, 
                            out.field$marginals.range.nominal[[1]])

# parameters for the SLOO process
ss <- 20 # sample size to process (number of SLOO runs)

# define the radius of the spatial buffer surrounding the removed point.
# Make sure it isn't bigger than 25
rad <- min(range.out, max(dist(coords)) / 4) 

alpha <- 0.05 # RMSE and MAE confidence intervals (1-alpha)


set.seed(199)

# run the function to compare both models
cv <- inlasloo(dataframe = dataframe, 
               long = 'long', lat = 'lat',
               y = 'cadmium', ss = ss, 
               rad = rad, 
               modform = list(modform, modform2),
               mesh = mesh, family = 'normal',
               mae = TRUE)
   


\end{example}

\section{Discussion and Conclusion}
 \label{section:conclusion}

The \textbf{INLAutils} package provides tools that allow users of \textbf{R-INLA} to visualise the results of their analysis more easily. We described the main graphical commands, including the \texttt{autoplot} command. Furthermore, we provide a concise coding framework to carry out and get an efficient visualization of the results of a spatial leave-one-out cross validation (SLOOCV) for spatial models fitted in \textbf{R-INLA} with the command \texttt{inlasloo}. 

Although our SLOOCV approach allows users to compare the predictive power of several models simultaneously, some limitations are inevitable. First, predictive scores based on the likelihood (as provided in \citet{LeRest2014} for example) are currently not included. Second, our approach only applies to purely spatial data, therefore spatio-temporal models are not considered. Future improvements of the package might include new graphical abilities and the incorporation of more models depending on the user's requests. 


\bibliography{lucas-etal-inlautils}-

\address{Tim Lucas\\
  University of Oxford, Li Ka Shing Centre for Health Information and Discovery, Big Data Institute\\
  Oxford OX3 7LF \\
  United Kingdom\\
  (ORCiD: 0000-0003-4694-8107)\\
  \email{timcdlucas@gmail.com}}

\address{Andre Python\\
  University of Oxford, Li Ka Shing Centre for Health Information and Discovery, Big Data Institute\\
Oxford OX3 7LF \\
United Kingdom\\
	(ORCiD: 0000-0001-8094-7226)\\
	\email{andre.python@bdi.ox.ac.uk}}

\address{David Redding\\
  University College London, Centre for Biodiversity and Environment Research\\
London WC1H 0AG \\
United Kingdom\\
	(ORCiD: 0000-0001-8615-1798)\\
	\email{dwredding@gmail.com}}

\end{article}

\end{document}